\documentclass[preprint]{acmart}

\usepackage{booktabs} 
\usepackage{rotating}
\usepackage{stfloats}
\setcopyright{rightsretained}

\newcommand{\beetweeness}{\beta}
\newcommand{\edge}{e}

\newcommand{\Edges}{\mathcal{E}}
\newcommand{\Nodes}{\mathcal{V}}
\newcommand{\safety}{s}
\newcommand{\risk}{\rho}
\newcommand{\location}{l}
\newcommand{\zscore}{z}
\newcommand{\coef}{\beta}
\newcommand{\standarderror}{SE}
\newcommand{\brier}{{\tt BS}}

\newcommand{\comm}[1]{}

\newcommand{\distance}{\textit{d}}
\newcommand{\speedlimit}{\textit{v}}
\newcommand{\width}{\textit{w}}

\newcommand{\hilliness}{\textit{h}}
\newcommand{\topology}{\theta}
\newcommand{\bikelane}{\textit{b}}

\acmDOI{10.475/123_4}

\acmISBN{123-4567-24-567/08/06}

\acmConference[KDD'19]{ACM KDD conference}{August 2019}{Anchorage, Alaska, US}
\acmYear{2019}
\copyrightyear{2016}

\acmArticle{4}
\acmPrice{15.00}

\begin{document}

\title{A Data-Driven Approach for Assessing Biking Safety in Cities}

\author{Sara Daraei}
\affiliation{%
  \institution{School of Computing \& Information \\ University of Pittsburgh}
}
\email{daraei.sara@pitt.edu}

\author{Konstantinos Pelechrinis}
\affiliation{%
  \institution{School of Computing \& Information \\ University of Pittsburgh}
}
\email{kpele@pitt.edu}

\author{Daniele Quercia}
\affiliation{%
  \institution{Bell Labs \\ Cambridge, UK}
}
\email{daniele.quercia@nokia-bell-labs.com}

\renewcommand{\shortauthors}{S. Daraei et al.}

\begin{abstract}
With the focus that cities around the world have put on sustainable transportation during the past few years, biking has become one of the foci for local governments  around the world. 
Cities all over the world invest in bike infrastructure, including bike lanes, bike parking racks, shared (dockless) bike systems etc. 
However, one of the critical factors in converting city-dwellers to (regular) bike users/commuters is safety. 
In this work, we utilize bike accident data from different cities to model the biking safety based on street-level (geographical and infrastructural) features. 
Our evaluations indicate that our model provides well-calibrated probabilities that accurately capture the risk of a biking accident. 
We further perform cross-city comparisons in order to explore whether there are universal features that relate to cycling safety. 
Finally, we discuss and showcase how our model can be utilized to explore ``what-if'' scenarios and facilitate policy decision making. 
\end{abstract}


%
%



\maketitle

\section{Introduction}
\label{sec:intro}

Transport engineers and urban planners have come to realize during the past few years that if we are to realize the vision of {\em smart cities}, that is, cities that are livable, sustainable and resilient, we need to move away from the car-centric mobility and turn to more sustainable modes of transportation. 
As a result cities have indeed turned to alternative modes of transportation, while promoting multi-modal mobility. Bicycles offer a promising transportation alternative to private vehicles, especially in areas with congestion, poor air quality, and high fuel prices \cite{klobucar2007network}.
Therefore, one of the major emphasis has been placed on urban biking as an alternative mode of transportation \cite{martens2004bicycle}. 
As a result during the past years the relevant infrastructure (e.g., bike lanes, bike parking racks, etc.) and services (e.g., shared bike systems) have enjoyed a significant growth \cite{bikes-growth}. 
This investment has resulted in an increased (perceived and/or actual) safety, either directly or through network effects \cite{jacobsen2003safety}, which consequently has a positive feedback on the propensity of people to ride a bicycle for their commute \cite{baker2009get,bikes-safety,bikes-safety2}. 

However, despite this increase in ridership the fraction of commuters that regularly use bikes is still relatively small as compared to other modes of transportation \cite{witlox2004evaluating}. 
It is thus, important to understand what factors are related with bike safety and how we can make riding safer in a city. 
While there exists literature that aims at understanding through surveys how people perceive biking safety\footnote{We discuss this and other relevant literature extensively in the following section.}, in this study we take a data-driven approach, using bike accident data from the cities of  London, Boston and Pittsburgh. 
We further collect street-level features from OpenStreetMap that capture both the geography as well as, the bike infrastructure on the streets of these cities. 
Using these data we build a model that quantifies biking safety as the relative probability of a {\em severe} biking accident in a given location. 
Our results indicate that the probability output of our models are well-calibrated both when evaluated (i) on a test set from the same city that the model was trained on, as well as, when evaluated (ii) on a dataset from a different city. 
The latter, i.e., a cross-city study/evaluation, is a piece that is missing in many studies in the area of urban computing and analytics, but at the same time is crucial in order to extract {\em universal} patterns (if any) of the phenomenon studied, which hold across cities. 

In brief, our results indicate that the presence of (protected) bike lanes is associated with a significant improvement in biking safety - as one might have expected. 
Furthermore, the speed limit, street topology (straight/curved), and the distance from an intersection are good predictors for the severity of a biking accident as well. 
While these results might not be surprising, and in fact they even might be expected from prior research or simply as plausible (and common sense) hypotheses, our work quantifies these relationships and enables their use in ``what-if'' scenarios. 
This can facilitate local government decision making as we elaborate on later. 
Thus, the main contributions of our work are: 

\begin{itemize}
\item We develop a bike safety model using bike accident and map datasets from 3 diverse cities. 
\item We perform a cross-city evaluation in order to identify possibly universal patterns related with biking safety. 
\item We also showcase how our models can be used by policy makers to evaluate the current infrastructure and decide on updates that maximize bike safety (possibly under other constraints).
\end{itemize}

The rest of the paper is structured as follows: 
Section \ref{sec:related} discusses related to our study literature and further differentiates our work. 
Section \ref{sec:data} describes the datasets used in our analysis as well as the biking safety model developed. 
We further provide our model evaluations in the same section. 
Section \ref{sec:app} shows how our model can be used for facilitating decisions on infrastructure updates that maximize biking safety under various constraints. 
Finally, Section \ref{sec:discussion} discusses the limitations of our study, while Section \ref{sec:conclusions} concludes our work. 

\section{Related Studies}
\label{sec:related}

In this section we will discuss relevant to our study research and position our study within this literature.  

For the past few decades, transportation planners have been trying to identify how an urban environment can attract more bike commuters. 
For example, Clarke \cite{clarke1992bicycle} explores the key factors in having a bike-friendly city. 
While Clarke focuses on the institutionalization of bicycle programs within agencies, safety is identified as an important factor that agencies will have to consider and overcome in order for biking to be fully integrated to the rest of the urban transport network. 
As a result there are several studies in the literature that aim at assessing the safety (or risk) of biking in the city and identifying characteristics of the street network that could help improve safety. 
A main differentiating point between these studies is the way the concept of safety is quantified. 
There are three ways that have been used to examine the concept of safety: 

\begin{itemize}
    \item {\bf Actual safety: }In this case the risk of biking is directly assessed through data that capture both accidents/fatalities as well as exposure of bicyclists to these risks (e.g., bike trip counts). 
    \item {\bf Perceived safety: }In this case risk of biking is assessed through user surveys, and hence, provide a subjective view of the road users. 
    \item {\bf Inferred safety: }When there is not a direct measure of quantifying the risk of biking, the latter can be inferred through indirect measures, such as the relative position or speed of motor vehicles to bicycles.
\end{itemize}

In order to calculate the actual safety, studies have relied on exposure estimates from national surveys (e.g., \cite{rodgers1995bicyclist}) or through video taping of traffic (e.g., \cite{harkey1998development,hunter1999study}). 
An interesting approach for dealing with the problem of estimating the exposure volume (e.g., the number of trips a bicyclist has taken, the number of bike trips over a specific segment etc.) has been presented in \cite{teschke2012route,harris2013comparing}. 
In these studies, the authors use a case-crossover approach, where data from bike accident locations are {\em matched} with a random point on the trajectory the bicyclist followed at the trip that led to the accident. 
The dataset that is created this way allows to  examine the influence of infrastructure on injury risk, while ensuring strict control for external  covariates (i.e., exposure to risk, cyclist traffic volume) and for personal and trip characteristics (e.g., propensity for risk-taking, time of day).

Given the possible lack of detailed data on bicycle accident or risk exposure, especially during the beginning of biking expansion in cities\footnote{In fact, even today that cities provide open information/data on bicycle accidents, these are constraint only to the accidents that are reported, which can be as low as 14\% \cite{gaarder1994bicycle}.}, researchers were also interested in assessing and quantifying the perceived safety of biking. 
Even though actual and perceived safety can differ, understanding how people perceive biking safety is particularly important; 
it can provide simple ways with which transportation and city authorities can {\em nudge} more people to use bicycle for their commute. 
To that end several studies have performed user surveys to understand what street features commuters associate with safety and/or which street aspects cyclist pay attention to (e.g., \cite{van2004cyclists,badgett1993bicycle,goldsmith1992reasons,aultman1998sidewalk,sharples1999use}). 

This research has led to the development of various indices that aim at quantifying the risk/safety associated with biking over a specific segment. 
In the late 1980s, early 1990s, the efforts were mainly focused on utilizing a combination of perceived and inferred safety.  
Davis \cite{davis1987bicycle} was the first to develop a score (Bicycle Safety Evaluation Index) with the objective of estimating the probability of an accident over a specific segment. 
The index combines information for the number of lanes, pavement, speed limit and similar variables. 
However, the weighting of each of these variables is {\em subjective} and it does not incorporate bicycle trips volume as an exposure variable. 
Despite its shortcomings the Davis work was the basis of the roadway condition index (RCI) and the Florida's Bicycle Coordinator segment condition index (SCI) \cite{epperson1994evaluating}. 
Landis \cite{landis1994bicycle} built on these theoretical models to develop the interaction hazard score (IHS), whose results are evaluated through surveys with bicyclists, essentially attempting to recreate the perceived safety. 

Many of these very early attempts to develop models for bike safety in a city had to deal with the absence of access to detailed data on bike accidents (and street level mapping) that we have today. 
Closest to our study is the work by Allen-Munley {\em et al.} \cite{allen2004logistic}, who used crash data from Jersey City to model the severity of a bicycle crash. 
In particular, the authors - similar to our study - build a logistic regression model for the severity of an accident using as independent variables various street features (e.g., speed limit, number of lanes etc.). 
The authors are focused more on a descriptive model, and hence, they do not evaluate the predictive power of the developed model, which is the main focus of our study, since this can drive educated recommendations for alterations in the infrastructure that potentially can improve biking safety. 
Furthermore, we also cross evaluate the models in different cities, in order to examine the presence of any {\em universal} patterns in terms of biking safety.

\section{Data and Methodology}
\label{sec:data}

In this section we will describe the datasets we used for our analysis as well as the biking safety model we developed. 

\subsection{Data}

{\bf Bike Accident Data: }
To perform our study we obtained data\footnote{Code and datasets are available at \url{https://github.com/daraei/BikeSafety-/tree/master/BikeSafety}.} on traffic accidents that involved bicycles from three metropolitan areas across Europe and North America. 
While these data are obtained from different sources (i.e., the Open Data portal of each city) they all provide the information we need in order to build the safety model. 
In particular, we collected data from: 

\begin{itemize}
    \item {\em London:} London's bike accident dataset is obtained from CycleStreet\footnote{\url{https:// bikedata.cyclestreets.net}} and covers the period between 2005 and 2017.
    \item {\em Boston:} Boston's dataset is obtained through Cambridge Open Data\footnote{\url{https://data.cambridgema.gov/Public-Safety/Crash-Map-2010-2013/4tke-xhvq}} and covers the period between 2010 and 2013.
    \item {\em Pittsburgh: } Pittsburgh's dataset is obtained through the Western Pennsylvania Regional Data Center\footnote{\url{https://data.wprdc.org}} and covers the period between 2004 and 2017.
\end{itemize}

\begin{table}[htbp]\centering

\begin{tabular}{l c c }

\toprule
\textbf{City } & \textbf{Years Covered} & \textbf{Sample Size}\\ 
\midrule
London    &      2005- 2017   &      3765   \\
Boston    &      2010- 2013   &      2157    \\
Pittsburgh    &      2004- 2017   &      1724   \\
\hline
\end{tabular}
\vspace{0.4cm}
\caption{Basic information for our bike accident data.}
\label{table:datasets}
\end{table}

Table \ref{table:datasets} provides some additional basic information from the data. 
While these datasets are compiled, curated and distributed by different entities, the subset of the information that we need for building a biking safety model is included in all three datasets. 
In particular, the tuple of interest is: {\tt <Lat, Lon, Accident Severity>}. 
While latitude and longitude is unambiguous information, the severity of an accident can be defined differently across the different cities. 
For example, in the London dataset the severity levels are specified as ``slight'', ``serious'' and ``fatal'' while in the Pittsburgh dataset they are categorized as ``not injures'', ``minor injury'', ``moderate injury'', ``major injury'', and ``killed''. 
Finally, in the Boston dataset the severity is represented by a Likert-line scale between 0 (no injury) to 4 (fatal).
In order to be able to build models that are comparable between the different cities, we have unified the labels for the severity of the accidents. 
In particular, we have used two labels, namely, (i) severe, and (ii) slight, and Table \ref{table:labels} shows the mapping between the lables in each city. 
We will further discuss the potential limitations from this in Section \ref{sec:discussion}. 

\begin{table}[ht]

\begin{tabular}{|ll|l|}
\hline
                                                  & Existing Labels    & Unified Labels           \\ \hline
\multicolumn{1}{|l|}{{London}}     & slight             & slight                   \\ \cline{2-3} 
\multicolumn{1}{|l|}{}                            & serious            & {serious} \\
\multicolumn{1}{|l|}{}                            & fatal              &                          \\ \hline
\multicolumn{1}{|l|}{{Boston}}     & 0 (no injury)      & slight                   \\
\multicolumn{1}{|l|}{}                            & 1 (light injury)   &                          \\ \cline{2-3} 
\multicolumn{1}{|l|}{}                            & 2 (fair injury)    & serious                  \\
\multicolumn{1}{|l|}{}                            & 3 (serious injury) &                          \\
\multicolumn{1}{|l|}{}                            & 4 (fatal)          &                          \\ \hline
\multicolumn{1}{|l|}{{Pittsburgh}} & not injures        & slight                   \\
\multicolumn{1}{|l|}{}                            & minor injury       &                          \\ \cline{2-3} 
\multicolumn{1}{|l|}{}                            & moderate injury    & serious                  \\
\multicolumn{1}{|l|}{}                            & major injury       &                          \\
\multicolumn{1}{|l|}{}                            & Killed             &                          \\ \hline
\end{tabular}
\vspace{0.5cm}
\caption{Label matching between datasets.}

\label{table:labels}
\end{table}

{\bf Street Features Data: } 
Having the latitude and longitude of a traffic accident allows us to obtain various information about the location that could be correlated with the severity of an accident involving a bicycle. 
Using OpenStreetMap's API we extract street-level features for each of the accident locations. 
These features include speed limit, the existence of a bikelane or not, the width and the length of the street segment, the hilliness, the topology of the street segment (i.e., whether it is straight or curved), as well as, the distance of the accident location from an intersection.

\textbf{Street Network: }
An important factor that we would like to examine as part of our model is the traffic volume on each street segment. 
However, we do not have access to this information, and hence, we take an indirect approach using as proxy appropriate network features. 
In particular, we extract from OpenStreetMaps the actual street network of each city, $\mathcal{N} = \{\Nodes, \Edges\}$.  
The set of nodes $\Nodes$ represents street intersections, while the set of edges $\Edges$ represents street segments connecting these intersections, i.e., edge $\edge_{i,j} \in \Edges$ exists if there is a street segment between intersection $i \in \Nodes$ and intersection $j \in \Nodes$. 
Using this network we calculated the edge  betweenness centrality for every edge $\edge \in \Edges$. 
The edge betweenness $\beetweeness_{\edge}$ of edge $\edge$ is simply the sum over all the possible pairs of nodes $(i,j)$, of the fraction of all pairs shortest paths between $i$ and $j$ that pass through edge $\edge$: 

\begin{equation}
    \beetweeness_{\edge} = \sum_{i,j\in \Nodes} \dfrac{\sigma_{ij}({\edge})}{\sigma_{ij}}
    \label{eq:between}
\end{equation}
where $\sigma_{ij}$ is the number of all shortest paths between nodes $i$ and $j$, while $\sigma_{ij}({\edge})$ is the number of those that pass through $\edge$.
The betweenness centrality of an edge $\edge$ is essentially proportional to the probability that $\edge$ will be part of one of the shortest paths between a randomly selected pair of nodes $i$ and $j$. 
Hence, one can easily argue that the betweenness centrality of a street segment in our network $\mathcal{N}$ is a good proxy for the overall traffic over the corresponding segment.

\begin{table*}[ht]
\centering
\setlength{\tabcolsep}{20pt}
\begin{tabular} {l r r r}
\toprule
\textbf{Variable} & \textbf{London} & \textbf{Boston} & \textbf{Pittsburgh}\\ 
\midrule
(Intercept)         &   4.372***    &  4.840***     &     2.979*** \\

speed-limit         &   11.071***  &   6.228***   & 12.332***  \\

Street width            &  8.937**      &   3.827*      &       6.382*** \\

betweenness         &  5.093***      &   4.827***    &       3.082**  \\

dist-intersect      &  -6.978***     &    -4.280**     &    -3.829**      \\

(hilliness) hilly   &  7.832***    &      7.374**       &       10.736***\\

(topology) curved       &   12.932***   &     6.482***       &       3.278***\\

(bikelane) with     &   -14.891***  &   -18.349***   &       -11.272***\\

\textbf{Interaction terms}     &          &       \\
speed-limit. betweenness     &   11.742**   &  4.729**   & 3.952*  \\
speed-limit. (bikelane) with     &   -21.922***   &  -23.749***  & -23.739***\\
speed-limit. dist-intersect &   4.281   &   4.839  & 3.398*\\

\bottomrule
\addlinespace[1ex]
\multicolumn{3}{l}{\textsuperscript{***}$p<0.01$, 
  \textsuperscript{**}$p<0.01$, 
  \textsuperscript{*}$p<0.05$}
\end{tabular}
\vspace{0.5cm}
\caption{Logistic regression coefficients for the risk model built with the datasets from different cities.}
\label{table:regression}
\end{table*}

\begin{figure*}[ht]
\centering
\includegraphics[scale=0.65]{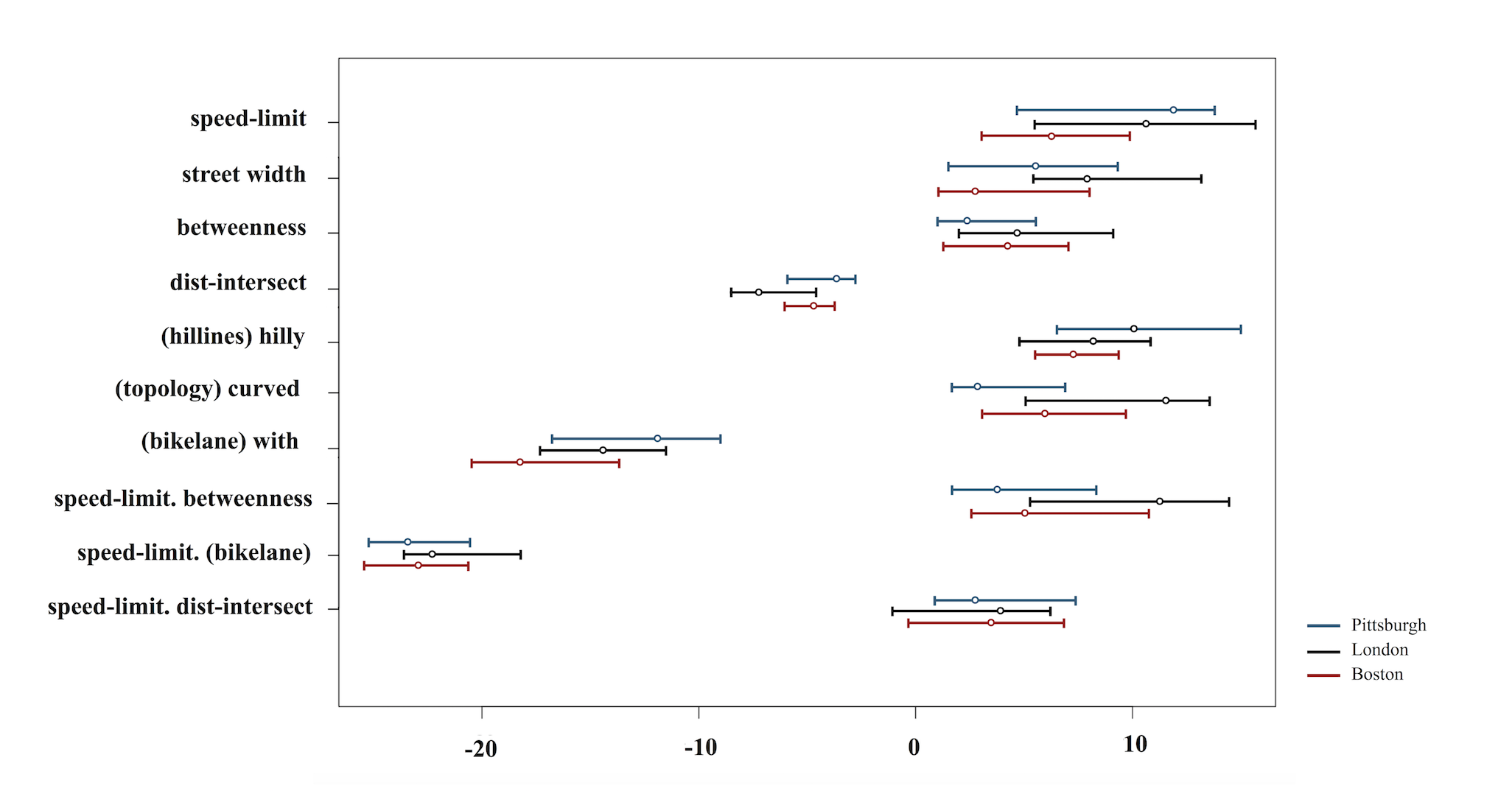}
\caption{The 95\% confidence intervals for the regression coefficients from the biking risk models built across different cities overlap for all independent variables.}
\label{fig:CI_models}

\end{figure*}

\subsection{Descriptive Models}
\label{sec:descriptive}

To reiterate, our goal is to build a model that will capture the safety $\safety$ (or risk $\risk$ depending on how it is viewed) of biking in an urban environment. 
Ideally, given data from bike accidents on a street segment a risk score would be assigned to the segment based on the number of these accidents. 
However, this number should be normalized with the total number of bike trips that go through this street segment. 
If we only know that there were 10 bike accidents over a street segment, this does not inform us about the biking risk $\risk$ associated with the segment. 
It can be anywhere from extremely low (e.g., if there have been 1 million bike trips over the segment), to completely risky (e.g., if there have been only 10 bike trips in total over the segment). 
We could then use this normalized accident count as our dependent variable for a safety model. 
Nevertheless, traffic data are not easily available/accessible, and this problem is particularly pronounced for bike traffic. 
Therefore, we rely on a slightly different definition of risk $\risk$ that has also been used in existing literature \cite{allen2004logistic}.  

In particular, we focus on a specific location $\location$ (rather than street segment), and define the risk associated with this location $\risk_{\location}$ as the probability of an accident that happens at $\location$ being {\em severe}. 
Simply put $\risk_{\location}$ is the conditional probability of an accident at $\location$ being severe given that an accident at $\location$ happened ($A_{\location}$, i.e.,  $\risk_{\location} = \Pr[\Sigma_{\location}|A_{\location}, \mathbf{x}]$. 
The aforementioned probability is also conditioned on the vector $\mathbf{x}$, which are the independent variables of our model as described in what follows. 
We model this probability through a logistic regression model, i.e.: 

\begin{equation}
    \Pr[\Sigma_{\location}|A_{\location}, \mathbf{x}] = \dfrac{1}{1+e^{-\mathbf{a}^T\cdot \mathbf{x}}}
    \label{eq:logistic}
\end{equation}

\begin{figure*}[ht]
\centering
\includegraphics[scale=0.6]{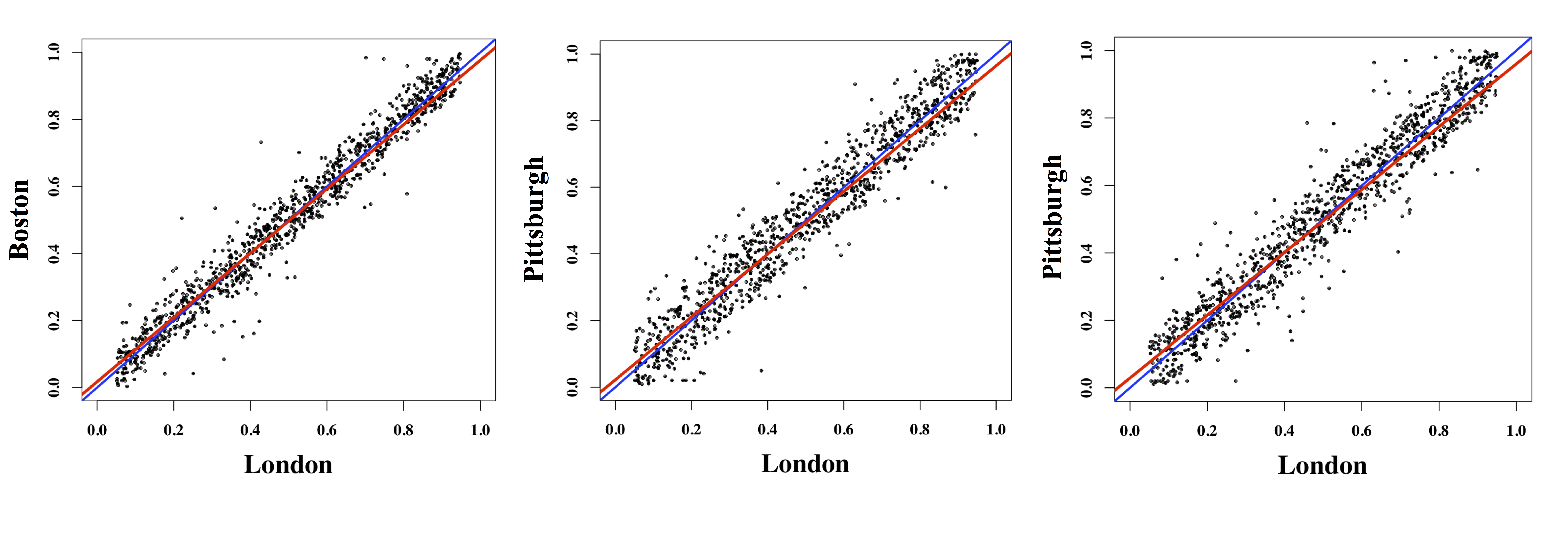}
\vspace{-0.4cm}
\caption{Severe accident probability for 1,000 random locations in San Francisco as estimated from all three different risk models. All models provide similar probabilities for the same location. }
\label{fig:multi_pred}
\end{figure*}

The features that we used as the independent variables of our model are: 

\begin{enumerate}
    \item {\bf Speed-limit }($\speedlimit$):   This variable captures the allowed maximum speed in each location. 
    
    \item {\bf Street width }($\width$):  This features represents the width of street in each location which also could be considered as an indicator for number of lanes.
 
    \item {\bf Distance to the intersection} ($\distance$): 
    This feature captures the distance to intersection for each accident. 

    \item {\bf Hilliness level }($\hilliness$): This variable including two categories of ``hilly'' and ``flat'' represents the steepness of each location.
    
    \item {\bf Street topology} ($\topology$):  This binary variable demonstrates if the accident happened in a curved or straight street. 
    
    \item {\bf Bikelane} ($\bikelane$): This binary variable captures the existence of bike-lane. 
    
    \item {\bf Street Betweenness} ($\beetweeness$):  This feature captures the edge betweenness centrality in the street network aforementioned. This variable can be thought of as a proxy for the traffic volume on the location of the accident.

\end{enumerate}

Table \ref{table:regression} presents the results obtained from each city. 
We train these models using the last 4 years that each dataset covers. 
The reason for restricting building the model with more (but older) data is twofold: {\bf (i)} the street features 15 years ago are most probably (very) different than what they are now, and {\bf (ii)} as people get more used to biking around the city and use better biking equipment that exists today the risks associated with it can change. 
Therefore, we want to restrict our model to only (fairly) recent data. 
As we can see all of the features are strongly correlated with the probability of a severe accident and for all three models the direction of these correlation is the same for the same independent variables. 
For example, an accident that happens on a central street (i.e., large betweenness) is expected to have higher risk for severity, while the presence of bike lanes reduces this risk.

However, apart from the direction of the correlations, we are also interested in how different the various coefficients of the models are. 
Hence, in Figure \ref{fig:CI_models} we present the 95\% confidence interval of the coefficients for each models. 
As we can see, even though the actual coefficients are different, their confidence intervals overlap, which essentially means that their is no statistical difference between the model coefficients\footnote{We also present in the appendix the z-scores for the difference of the model coefficients.}. 
In other words, biking safety appears to be correlated with street features in similar ways across the different cities we examined.

\begin{figure*}[ht]
\centering
\includegraphics[scale=0.65]{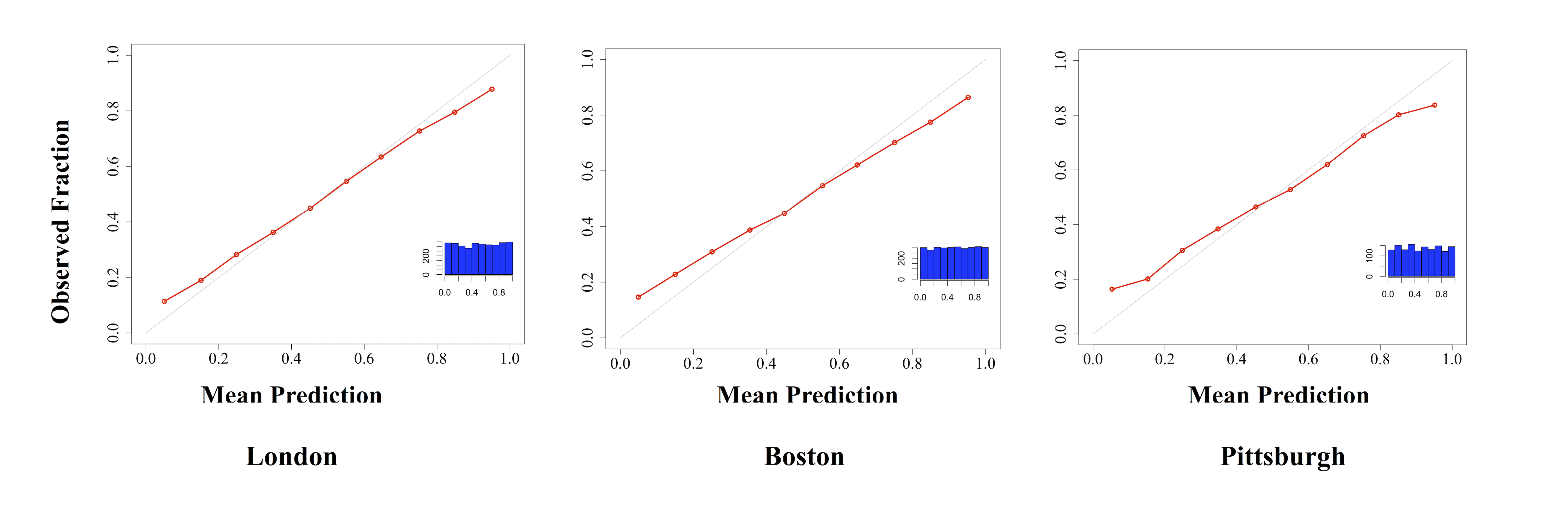}
\vspace{-0.4cm}
\caption{ Reliability curves of the biking risk model probabilistic predictions.}
\label{fig:calibration}
\end{figure*}

Finally, we selected a 1,000 random locations from a different city (San Francisco, California) and apply our different risk models to estimate the conditional probability of a severe bike accident on that location. 
Our goal with this experiment is to compare the differences in the predictions provided by the various models for the same points. 
Figure \ref{fig:multi_pred} presents the scatter plots of these predictions, where the two axes presents the risk probability based on the models from two different cities. 
As we can see the points follow closely the $y=x$ line, which means that all models give similar risk probabilities for the same locations.

\subsection{Evaluating the Models' Predictive Power}
\label{sec:predictive}

Apart from the descriptive nature of the biking risk models explored above, we are also interested in their predictive power. 
To evaluate the predictive power of each model we split each dataset to a training and a testing set. In particular, for each city we use the first two years in our dataset for training and the last two as our out-of-sample testing set. 
Table \ref{tab:results2} present the out-of-sample accuracy for each model, as well as, the Brier score. 
The latter is a measure of probability calibration. 
Specifically, in a probabilistic model, its classification accuracy paints only part of the picture.  
For example, two models $M_1$ and $M_2$ that both predict an accident will be severe they will have the same accuracy. 
However, if $\Pr_{M_1}[\Sigma_{\location}|A_{\location}, \mathbf{x}] = 0.9$ and $\Pr_{M_2}[\Sigma_{\location}|A_{\location}, \mathbf{x}] = 0.55$, the two models have different probability calibration. 
Brier score \cite{brier1950verification} quantifies this calibration for each model. 
In particular, for the case of binary probabilistic prediction, the Brier score is calculated as:

\begin{equation}
\brier = \dfrac{1}{N}\sum_{i=1}^N (\pi_i-y_i)^2
\label{eq:brier}
\end{equation}
where $N$ is the number of observations, $\pi_i$ is the probability assigned to instance $i$ of being equal to 1 and $y_i$ is the actual (binary) value of instance $i$.  
The Brier score takes values between 0 and 1 and as alluded to above evaluates the calibration of these probabilities, that is, the level of confidence they provide. 
The lower the value of $\brier$ the better calibrated the output probabilities are. 
Continuing on the example above a 0.9 probability is better calibrated compared to a 0.55 probability (when the ground truth is label 1) and hence, even though $M_1$ and $M_2$ have the same accuracy, $M_1$ is better calibrated (lower Brier score -- 0.01 compared to 0.2025).  
As we can see all models exhibit out-of-sample accuracy higher than 82\%, while their calibration is very good as captured by the Brier score. 
In particular, the Brier score for each model is much lower as compared to the corresponding score for a climatology model. 
The latter is a typical baseline Brier score used for evaluating the quality of a model, and is obtained through a reference model that assigns to each data point the base probability of class 1. 
Using the Brier score of the climatology model  $BS_{ref}$, we can further calculate the \textit{skill} of probability estimates using the Brier Skill Score ($BSS$) 
Specifically, if $BS_{\text{ref}}$ is the Brier score of the climatology model, then $BSS$ is calculated as:  
\begin{equation}
    BSS = 1 - \frac{BS}{BS_{\text{ref}}} 
    \label{eqn:brier_skill}
\end{equation}
$BSS$ will be equal to 1 for a model with perfect calibration, i.e., $BS = 0$. 
A model with no {\em skill} over the climatology model, will have a value of 0 since $BS = BS_{\text{ref}}$. 
If $BSS < 0$, then the model exhibits less skill than even the reference model.

\begin{table}[htbp]\centering
\begin{tabular}{l c c c c}

\toprule
\textbf{City } & \textbf{Accuracy} & \textbf{BS} & \textbf{BS-Baseline} & \textbf{BSS}\\ 
\midrule
London    &      0.885   &      0.119   & 0.14 & 0.15\\
Boston    &      0.832   &      0.169  & 0.19 & 0.11 \\
Pittsburgh    &      0.821   &      0.178 & 0.21 & 0.15 \\
\hline
\end{tabular}
\vspace{0.5cm}
\caption{Out-of-sample accuracy and Brier score for the different models. }
\label{tab:results2}
\end{table}

Finally, we evaluate the accuracy of the probability output of each model by deriving the probability calibration curves for the test set in each case. 
In order to compute the accuracy of the predicted probabilities we would ideally want to have several bike accidents (e.g., 100) happen at the same location $\location$. 
If the model assigned a 75\% probability of an accident at location $\location$ being severe,  then we would expect about 75 of the accidents observed in $\location$ to be severe.  
However, this is clearly not realistic to have (at least for the vest majority of the locations) and hence, in order to evaluate the accuracy of the probabilities we will use all the accidents in our dataset. 
In particular, if the predicted probabilities were accurate, when considering all the accidents where an accident was predicted to be severe with a probability of x\%, then a severe accident should have been observed in (approximately) x\% of these accidents. 
Given the continuous nature of the probabilities we quantize them into groups that cover a 10\% probability range. 
Figure \ref{fig:calibration} presents the predicted probability of a severe accident on the x-axis, while the y-axis presents how many of these accidents were indeed severe. 
Furthermore, we present in the inset figure for each calibration curve, the distribution of the predictive probabilities, which as we observe cover fairly uniformly the whole range of probabilities. 
As we can see the calibration curve is very close to the $y=x$ line, which practically means that the predicted probabilities capture fairly well the actual biking risk  probabilities. 




\subsection{Cross-City Models}
\label{sec:cross}

In Section \ref{sec:descriptive}, we compared the coefficients for the safety models we built for the different cities. 
Here we want to evaluate the ability of a model trained on a specific city to predict the severity of accidents in a different city. 
These cross-city evaluations and comparisons are largely absent from the current literature in urban informatics, but are very important. 
The ability to predict the severity of accidents in city A based on a model that was trained with data from city B, is a good indicator that there might be universal patterns in biking safety. 
Simply put, knowledge obtained from a specific city might be transferable (and generalizable) to another city. 

\begin{figure*}[ht]
\centering
\includegraphics[scale=0.65]{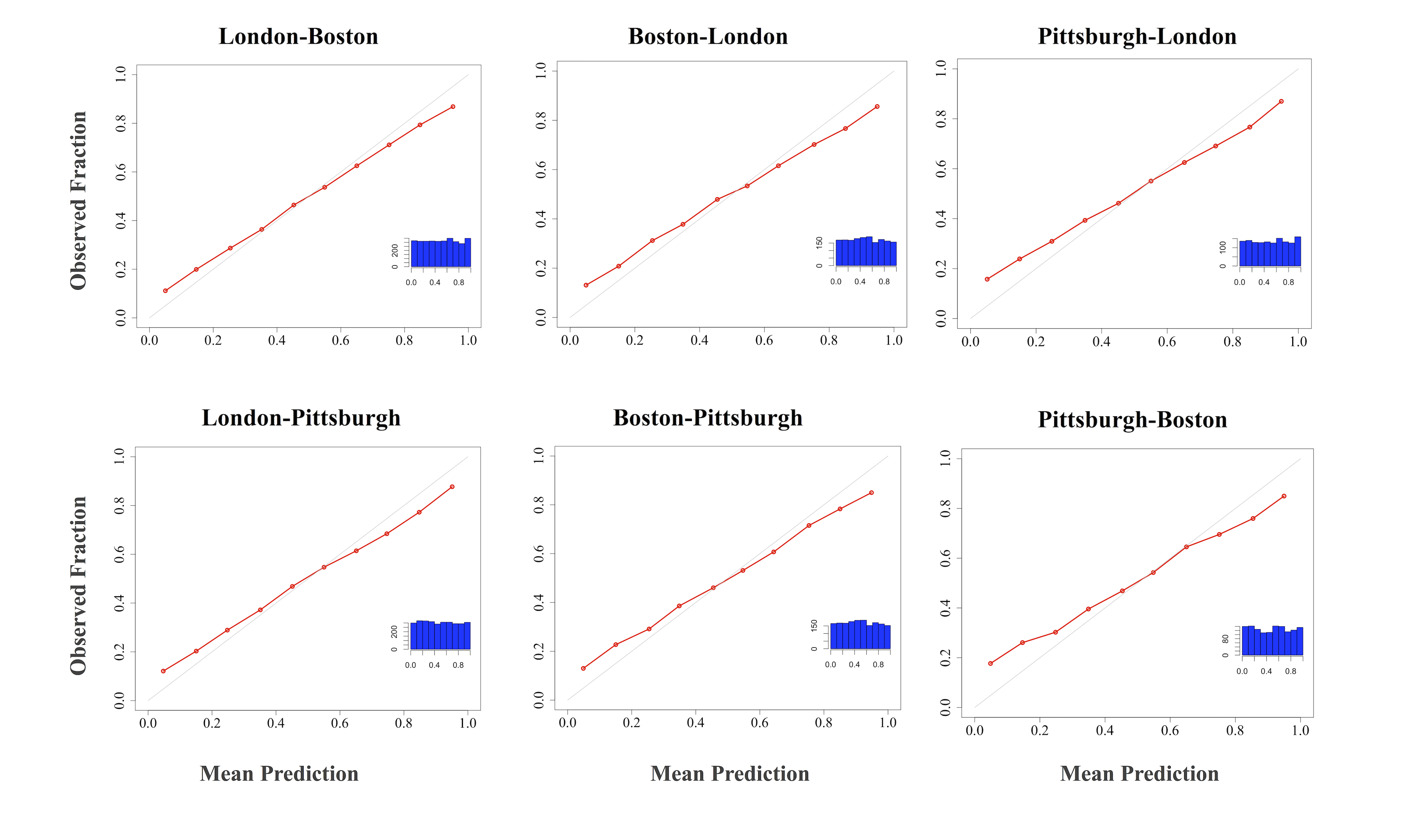}
\vspace{-0.5cm}
\caption{Calibration curves for out-of-{\em city} severe biking accident predictions. }
\label{fig:paircalibration}
\end{figure*}

In particular, we use the model trained with data from each city in our study to predict the severity of accidents in the other two cities in our study (6 pairs in total). 
Table \ref{tab:cross_city_results} presents the accuracy and the Brier score for these experiments, while Figure \ref{fig:paircalibration} shows the reliability curves for these predictions. 
As we can see, the accuracy, while reduced as compared to the out-of-sample performance on the same city, is still high. 
The same is true for the Brier score (and the corresponding Brier skill scores). 
In particular, the performance of a model trained on London data and evaluated on Boston data - and vice versa - exhibit almost as good performance as the within city evaluations. 
The dataset from the city of Pittsburgh is much smaller and this could impact the model quality, especially when used in this cross-city experiment. 
Finally, Figure \ref{fig:paircalibration} presents the calibration curves for the different pairs of training-test cities. 
As we can see again, the calibration of the probabilities are good and the curve is close to the $y=x$ line.

\begin{table*}[ht]
\centering
\setlength{\tabcolsep}{18pt}
\begin{tabular}{l l c c c c}

\toprule
\textbf{Training City } & \textbf{Testing City } & \textbf{Accuracy} & \textbf{BS}& \textbf{BS-Baseline} & \textbf{BSS}\\ 
\midrule
London    &  Boston  &  0.857   &      0.142  & 0.17 & 0.16 \\
London    &  Pittsburgh  &  0.822   &      0.177   & 0.21 & 0.16\\
Boston    &   London  &  0.806   &       0.193    & 0.22 & 0.12\\
Boston    &   Pittsburgh  & 0.797   &      0.222    & 0.24 & 0.08\\
Pittsburgh    &  London   & 0.786   &      0.210 & 0.24 & 0.13\\
Pittsburgh    &  Boston   & 0.763   &      0.236 & 0.25 & 0.06\\
\hline
\end{tabular}
\vspace{0.4cm}
\caption{Accuracy and Brier score for out-of-{\em city} evaluations. }
\label{tab:cross_city_results}
\end{table*}

\section{Case Study: Improving Bike Safety Under Constraints}

\begin{figure}[ht]
\centering
\includegraphics[scale=0.25]{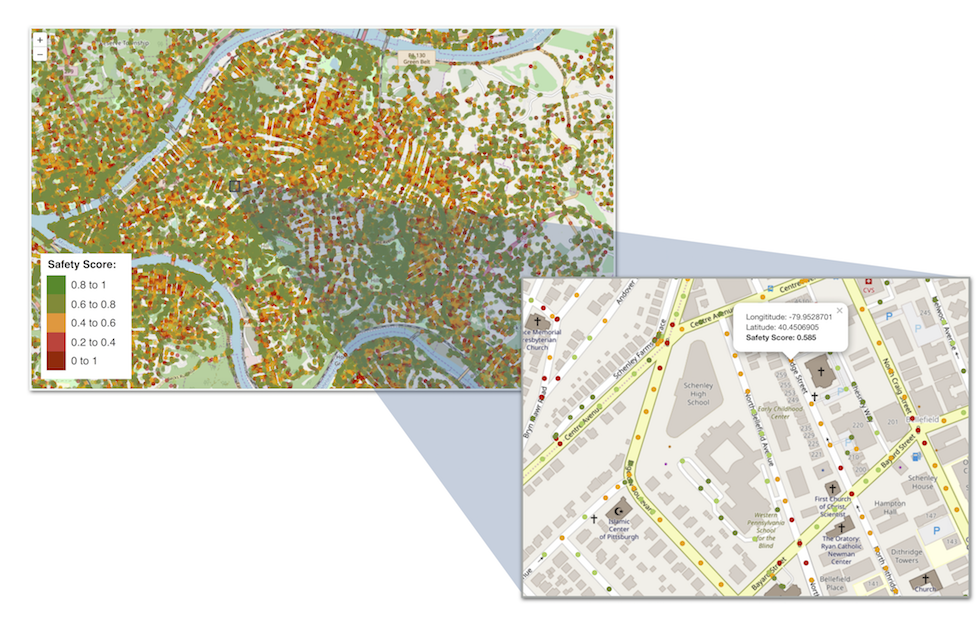}
\caption{Biking safety in the city of Pittsburgh as calculated from our model. }
\label{fig:Pitt_map}
\end{figure}

This type of models allow us to evaluate the biking safety within a city and can facilitate policy decisions with respect to infrastructure updates. 
For example, Figure \ref{fig:Pitt_map} presents the interactive map of results we obtained from our model for the whole city of Pittsburgh. By clicking on each point, users are able to be aware of the its safety score. 
In particular, for every point on the street network we estimate the safety $\safety$ (which is $1-\risk$) associated with this point and visualize it on the map. 
Safe locations (i.e., low risk of severe accident) are colored green, while risky locations are colored red. 
Using these results one can start identifying locations that are not as safe as needed and explore what infrastructure changes (e.g., widening of a street, installation of a bike lane etc.) are going to have the maximum impact on biking safety, possibly under budgetary (and policy) constraints.

\begin{figure}[ht]
\centering
\includegraphics[scale=0.5]{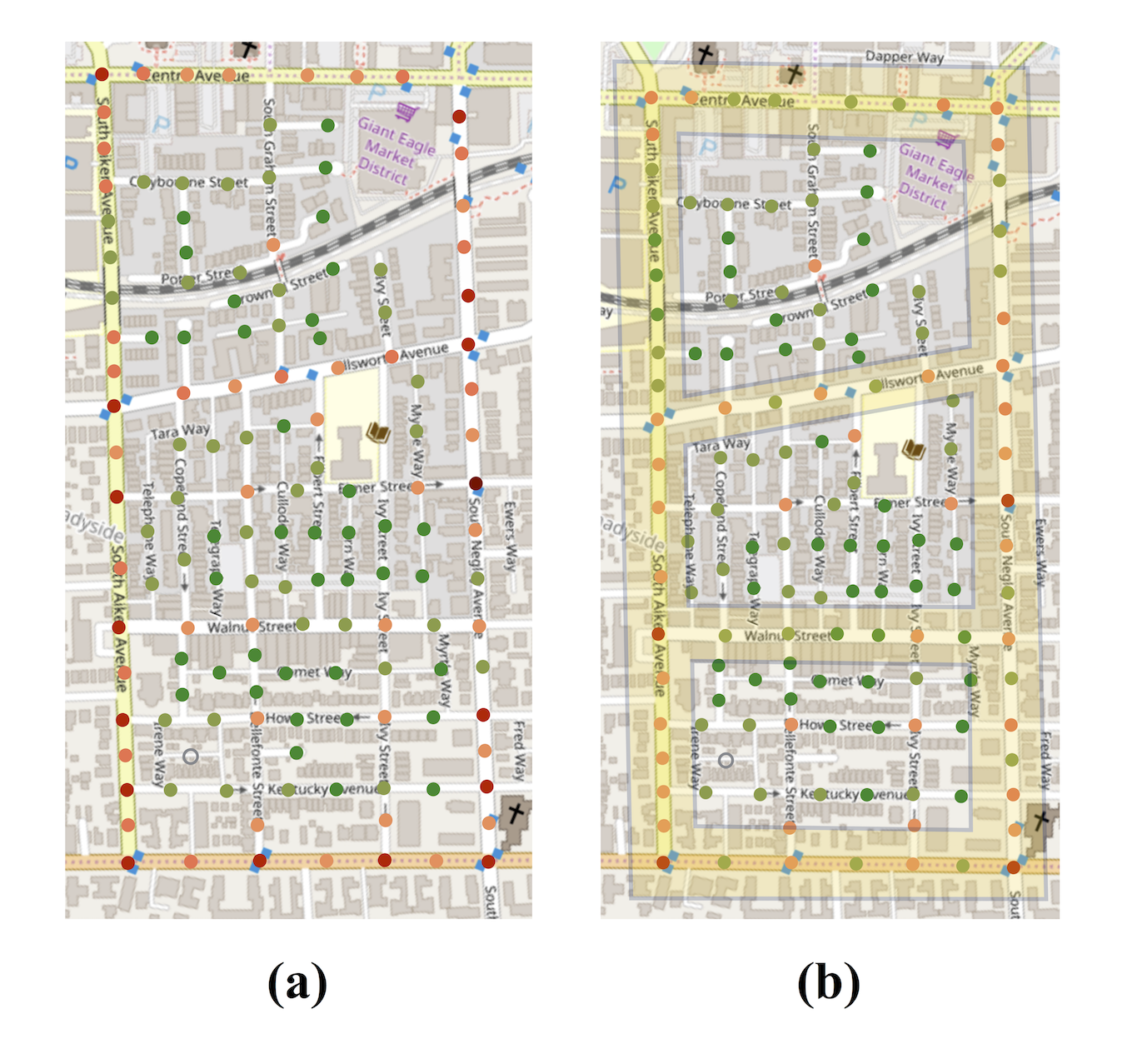}.
\caption{The examined area has an average safety score of 0.54. By adding bike lanes to the main streets, the safety score increases to 0.68; a 26\% increase in the average safety score.}
\label{fig:Pitt_map2}
\end{figure}

For example, Figure \ref{fig:Pitt_map2} (a) presents an 2.8 square miles area located within the city of Pittsburgh. The selected area, currently, does not include separate bikelanes for cyclists' transportation. The average safety score obtained for this area is $\safety = 0.54$. 

Let us assume that we would like to enhance biking safety for this area and we have budget for adding bike lanes. Furthermore, we have the constraint that we cannot install bike lanes in local streets (which typically have low traffic anyway). 
Figure \ref{fig:Pitt_map2} (b) presents the safety scores for the same area but now assuming that bike lanes have been installed in all of the non-local streets. 
Now the average safety score has increased to $\safety = 0.68$. 
This corresponds to an approximately 26\% increase in the average biking safety within the area by {\em simply} adding bike lanes in non-local streets. 

Obviously, with cities being cash-strapped today, they cannot just install bike lanes, or widen all the streets. 
However, models similar to the ones we developed in this study, allow policy makers to better understand the correlations between the characteristics of the existing infrastructure and how one could potentially improve the safety for bikers. 
In fact, one could explore a variety of (viable) options and decide which one is the {\em optimal} based on the criteria of interest (e.g., trade off between safety and cost etc.).

\label{sec:app}

\section{Discussion and Limitations}

\label{sec:discussion}

Our study contributes to the literature of urban informatics and more specifically to the area of computational transportation modeling. 
While models for biking safety have been developed in the past they are typically based on a small number of observations and most importantly cover a single city/area. 
In our study we aim at identifying aspects of biking safety that are similar across cities. 
However, despite the ability of our models to provide reliable predictions across different cities, we cannot really conclude universality of the results, but rather only within the set of cities we examined.

Furthermore, we collected the street level data for our models from OpenStreetMap. 
While OpenStreetMap is expected to provide high quality information in big cities like London, Boston and Pittsburgh, it is still a crowdsourcing platform and hence, it is possible to still have incomplete/wrong information. 
However, the general trends identified are not expected to be significantly impacted by the crowdsourcing nature of OpenStreetMap. 
One of the possible limitations of our study is that some of the street features we collected from OpenStreetMap are contemporary and they might have been different when the actual accidents data were recorded. 
For example, the speed limit, or the presence of a bikelane might have changed over the years. 
We have tried to account for that by analyzing only recent data. 
Moreover, the original severity labels for an accident have different categories. 
Although we unified them as demonstrated in Table \ref{table:labels}, there is still a subjective element on how the original data were classified. 

Finally, other classification models (e.g., tree classifiers, RNNs etc.) could possible perform better in predicting the severity of a bike accident. 
However, we wanted a model that is easily interpretable. 
This is particularly important for policy makers that want to make decisions based on the information provided by the model. 
At the same time more independent variables, e.g., features obtained from visual cues - presence of trees, head-in parking, presence of billboards etc. - could provide further information and further improve the quality of the model. 
We plan to explore similar features in future studies.

\section{Conclusions}
\label{sec:conclusions}

In this work our objective is to build an appropriate model that will allow us to understand the relationship between street infrastructure features and biking safety. 
To achieve our objective we obtain traffic accident data which involved a bicycle from three metropolitan areas, namely, London, Boston, and Pittsburgh. 
We further collected street features for the locations of the accidents and built a logistic regression model for each city for the probability of a severe accident, conditional to the presence of an accident. 
Our results indicate that these models are transferable between cities (at least the ones we examined). 
In particular, the coefficients are statistically the same in all three models, while each model has very good predictive power for accidents that happened in a different city. 
Finally, we show how our models can be useful to policy makers by evaluating the current state of bike safety within a city and analyzing ``what-if'' scenarios for infrastructure updates.

\bibliographystyle{ACM-Reference-Format}
\bibliography{sample-bibliography}

\appendix

\section*{Appendix} 
\subsection*{A. z-scores for the difference of the regression coefficients of the risk models}
For completeness, we also calculated the z-score for the difference of the coefficients between pair models trained from data obtained from different cities. 
In order to compare two regression coefficients $\beta_1$ and $\beta_2$ for the same independent variable obtained through models $M_1$ and $M_2$ we can calculate the z-score for their difference as \cite{clogg1995statistical,paternoster1998using}: 

\begin{equation}
    \zscore =  
    \dfrac{\coef_{1} - \coef_{2}}{\sqrt{\standarderror\coef_{1}^2+\standarderror\coef_{2}^2}}
    \label{eq:z-score}
\end{equation}
where $SE\beta$ is the standard error of coefficient $\beta$. 
This z-score is the statistic we can use to perform the following hypothesis test:

\begin{eqnarray}
H_0: & \beta_1 = \beta_2 \\
H_1: & \beta_1 \neq \beta_2 
\label{eq:hyp}
\end{eqnarray}

Table \ref{table:comparison} presents the difference in the corresponding regression coefficients and the corresponding p-value. As we can see none of the differences is  statistically significant (at the typical 5\% level) which means that we cannot reject the null hypothesis that the two coefficients are statistically the same. 
This translates to the independent variable having similar effects in the different cities.

\begin{table*}[hbt!]
\centering

\begin{tabular} {l r  r  r }
\toprule
\textbf{Variable} & \textbf{London vs. Boston}  & \textbf{London vs Pittsburgh} & \textbf{Boston vs. Pittsburgh}  \\ 
\midrule
(Intercept)         &   -0.468    (.471) &  1.393  (.373) &    1.861 (.251) \\

speed-limit         &   4.843 (.103) &   -1.261  (.284)& -6.101 (.061)  \\

Street width            &  5.11 (.126)    &   2.555 (.275)      &       -2.555 (.101)\\

betweenness         &  0.266  (.201)    &   2.011 (.052)  &       1.745 (.071) \\

dist-intersect      &  -2.698  (.275)    &    -3.149 (.307)     &    -0.451 (.065)      \\

(hilliness) hilly   &  0.458 (.347)    &     -2.904 ( .141)      &      -3.362 (.094)\\

(topology) curved       &   6.451 (.171)   &     9.654  (.234)     &       3.542 (.053) \\

(bikelane) with     &   3.458 (.289) &   3.619   (.081) &      -7.077 (.242)\\

\textbf{Interaction terms}     &          &       \\
speed-limit. betweenness     &   7.013 (.101) &  7.791   (.065)  &0.777 (.053) \\
speed-limit. (bikelane) with     &   1.827 (.242) &  -0.01 (.218) & -0.23 (.818)\\
speed-limit. dist-intersect &   -0.558 (.051) &   0.883 (.053) & 1.441 (.068)\\

\bottomrule
\addlinespace[1ex]

\end{tabular}

\caption{Difference in regression coefficients and corresponding p-value }
\label{table:comparison}
\end{table*}

\end{document}